\documentclass[12pt]{article}

\usepackage{amsmath,amsfonts,amssymb,latexsym}

\newtheorem{theorem}{Theorem}[section]

\newtheorem{definition}{Definition}[section]
\newtheorem{remark}{Remark}[section]

\numberwithin{equation}{section}
\def\be{\begin{equation}}
\def\ee{\end{equation}}
\def\bq{\begin{eqnarray}}
\def\eq{\end{eqnarray}}
\def\beq{\begin{eqnarray*}}
\def\eeq{\end{eqnarray*}}
\newcommand{\GA}{\alpha}

\newcommand{\GD}{\delta}

\newcommand{\GK}{\kappa}

\newcommand{\GR}{\rho}

\begin{document}
\begin{center}
{\huge Talking About Singularities}

\vspace{1cm}

{\large Spiros Cotsakis}\\

\vspace{0.5cm}

{\normalsize {\em Research Group of Geometry, Dynamical Systems and
Cosmology}}
\\ {\normalsize {\em Department of Information and
Communication Systems Engineering}}
\\ {\normalsize {\em University
of the Aegean}}
\\ {\normalsize {\em Karlovassi 83 200, Samos,
Greece}}
\\ {\normalsize {\em E-mail:} \texttt{skot@aegean.gr}}
\end{center}

\begin{abstract}
\noindent We discuss some aspects of recent research as well as more general
issues about motivation, useful methods and open problems in the
field of cosmological singularities. In particular, we review some
of the approaches to the general area and  include  discussions of
the method of asymptotic splittings, singularity and completeness
theorems and the use of the Bel-Robinson energy to prove
completeness theorems and classify cosmological singularities.
\end{abstract}

\begin{quote}
`Knowledge should be shared', yes, of course it should,  but it is
extremely doubtful whether an organic chemist would put missionary
zeal into an attempt to interest the pure mathematician in his
latest researches, or vice versa. To be realistic, the sharing of
knowledge throughout the whole field of science is a pious hope, not
a passion, and it is only passion that could overcome the
inescapable difficulties of communication. {\em J. L. Synge}.
\end{quote}

\section{Introduction}
Singularities have always been important, and all successful
theories of physics have them by necessity. From the Newtonian
dynamics of particles and fields, to relativistic problems, to
quantum dynamics, the ever-presence of singularities can for the
most part be adequately treated one way or another, even avoided or
ignored, although in some cases their consideration has triggered
fundamental changes.

But it is different in cosmology. Here, even though there are areas
where the consideration of singularities can for the most part or
wholly be neglected, no fundamental discussion of the basic
cosmological issues can go on without facing them sooner or
(preferably!) later. Up to the recent past, discussions of
cosmological singularities have revolved around the same
philosophical issues that we all know and grew up to respect, but
which after a while any young, inquiring and productive cosmologist
would wish to by-pass. But with certain exceptions. Not all
discussions about singularities in cosmology are philosophical, they
can be very mathematical indeed and at the same time touch upon very
basic physical issues, probably at the cost of creating a situation
that many people  (even the creators themselves!) would wish to
overpass.

We believe that there lies ahead of us a golden era for research in
cosmological singularities. The basic pillars are now almost done,
the proliferation of newer and more interesting cosmological models
is a call to arms, the mathematical techniques are better understood
and explained and are definitely there waiting to be used, and
cosmological observations, although richer than ever, do not
constrain the possible models to any considerable degree. It is
unfortunate that all this richness is often stated as a source for
despair, it is a paradise for those who don't expect ready-made
answers.

In this review we are set to present the subject of cosmological
singularities from a new angle, that of a logical and simplest form
which however, may not be the most psychologically natural one. For
motivation purposes, discussions of a new but already interesting
subject are perhaps equally important to precise definitions and
results, at least for the novice, and we have tried to present both
here (resulting unfortunately in the usual unlucky compromise).
 In the next section after a short review of the basic definitions of completeness and
singularity, we introduce the problem of cosmological singularities
in two stages: Firstly, it is set at the level of special, exact
cosmologies that is when we have chosen a particular spacetime, a
specific matter component and a geometric action to describe the
gravitational field, for instance a fluid-filled Friedmann universe
in general relativity. Secondly, we formulate the singularity issue
at the different level of leaving the choice of spacetime
unspecified. This leads us in turn into two ways of attack, one
through spacetime differential geometry to formulate and prove
general results about geodesic incompleteness (singularity
theorems), the other being the analysis of the field partial
differential equations describing the evolution of Einsteinian
spacetimes.

It is important to realize that any information about the
singularities of a system in the general case (that is without
assuming any symmetry) cannot be usefully reduced to give us
information about particular instances of the field in situations of
symmetry (and vice versa of course!). This last case requires a
different analysis. A concrete general technique, the method of
asymptotic splittings, designed specifically to deal with the
approach to a spacetime singularity of exact cosmologies is
presented in section 3. The aim of this method is to decompose the
vector field defining the model in a suitable way and follow the
dominant features of it all the way down the finite time
singularity. This leads to the construction of asymptotic series
expansions of the unknowns in the neighborhood of the singularity,
from which valuable information about the field can be extracted.
Some examples of the application of the method of asymptotic
splittings are discussed. This method is an adaptation to real
dynamical systems (finite dimensions) of the philosophy of certain
blow-up techniques in partial differential equations, and of methods
that exist in the field of complex differential equations
(singularity analysis).

Next we move on, in section 4, to discuss sufficient and necessary
conditions for the occurrence of singularities. This section,
besides the classic singularity theorems, includes more recent
results about singularities associated primarily with the name of Y.
Choquet-Bruhat. We also provide applications of these criteria for
singularity formation to the isotropic category and show how these
lead to a novel classification of spacetime singularity types
therein through the use of the Bel-Robinson energy. We also use the
Bel-Robinson energy to give a new characterization of the
g-completeness and eternal acceleration of Robertson-Walker (RW)
spacetimes.

The absence of a discussion section at the end is a healthy one, and
conclusions are out of place at this initial stage of development of
this most beautiful subject.

 I would like to thank many colleagues and
collaborators, especially Ignatios Antoniadis, John Barrow, Yvonne
Choquet-Bruhat, Ifigeneia Klaoudatou, John Miritzis, Vincent
Moncrief, Alan Rendall and Antonios Tsokaros, whose work and ideas I
have freely used  in this review, for many stimulating discussions
about various important issues associated with spacetime
singularities. Part of the research reviewed here was conducted
while the author visited CERN-Theory Division in Geneva and IHES in
Bures sur Yvette and I thank those institutions for allowing me to
use their excellent facilities.

\section{Approaches to the singularity problem}
In this section we give the basic definitions of what it means for a
spacetime to be singular. We then discuss the general problem of
spacetime singularities in the framework of exact cosmological
solutions (physical approach). Finally, we extend the formulation of
the problem to the case of spacetimes without symmetries and discuss
various more mathematical approaches used to tackle this most basic
cosmological issue.

\subsection{Definitions}
We give in this subsection the basic definitions of complete as well
as singular spacetimes. Let $(\mathcal{M},g)$ be a spacetime with
$\mathcal{M}=(a,b)\times\Sigma$, where $(a,b)$ is an interval in
$\mathbb{R}$ (possibly the whole line), and  the spatial slices
$\Sigma_{t}\,(=\Sigma\times \{t\})$ are spacelike submanifolds
endowed with the time-dependent spatial metric $g_{t}$. The standard
geometric definition of  spacetime completeness in general
relativity is the following.
\begin{definition}[Geometric]
We say  that the spacetime  $(\mathcal{M},g)$ is non singular if it
is causally geodesically complete, that is every causal (i.e.,
timelike or null) geodesic $\gamma :
(a,b)\subset\mathbb{R}\rightarrow\mathcal{M}$, defined in a finite
subinterval, can be extended to the whole real line.
\end{definition}
However, this definition is not useful in evolution problems because
in such studies one is typically interested not in the behaviour of
geodesics but in that of the \emph{dynamical} variables of the
problem. These satisfy differential equations such as the Einstein
or some other field equations. In such problems one is typically
interested in proving local or global in time existence of the
spacetime in question. This leads us to the following alternative.
\begin{definition}[Dynamical]
We say  that the spacetime  $(\mathcal{M},g)$  exists globally  as
a solution of the field equations if
$$||(\mathbf{g}_{t},\partial\mathbf{g}_{t},
\mathbf{\Theta})||_{L^p}<\infty,\quad\textrm{uniformly for all}
\,\,p,\,t .$$
\end{definition}
This definition, deliberatively vague, means intuitively that a
spacetime such that its spatial metric
 variables $\mathbf{g}_{t}$, their derivatives $\partial\mathbf{g}_{t}$ and any matter fields
$\mathbf{\Theta}$ present are finite for all time, must be
considered as globally existent. In a specific problem of course one
must choose the appropriate norms so that they make sense with the
given conditions on the various data. However, even when, as it is
usually the case, global existence comes together with non
singularity (in the above sense of geodesic completeness), one must
prove that this is indeed the case as the two notions are different.

The negations of the above definitions lead to the notions of
singular and finitely existent spacetimes.
\begin{definition}[Geometric]
We say that $(\mathcal{M},g)$ is singular if it is geodesically
incomplete, that is there is at least one incomplete causal geodesic
$\gamma : (a,b)\subset\mathbb{R}\rightarrow\mathcal{M}$.
\end{definition}
For a spacetime to have only  a finite duration, we require the
following
\begin{definition}[Dynamical]
We say that $(\mathcal{M},g)$  exists as a solution of the field
equations for only a finite time if the following holds:
$$||(\mathbf{g}_t,\partial\mathbf{g}_{t},\mathbf{\Theta})||_{L^p}\quad
\textrm{becomes unbounded at some}\,\, p, t. $$
\end{definition}
These definitions allow us to make several remarks.
\begin{remark}[Equivalence]
The two definitions, geometric and dynamical, of completeness on the
one hand, and singularity on the other, are considered as
equivalent, but there is to our knowledge no \emph{formal} proof of
this in the literature. So one must check for instance that a
spacetime that exists globally for all future time is also
geodesically complete to prove that it contains no singularities.
For an example of this, see \cite{cbm02,cbc}.
\end{remark}

\begin{remark}[True singularity]
It is important to emphasize that one should be careful not to mix a
singularity in the congruence of a bunch of geodesics with a true
spacetime singularity. The singularity theorems (see below) give
conditions  when one ends up with a true spacetime singularity, the
proof, however, of such statements is by contradiction and therefore
one cannot use such arguments to really \emph{construct} incomplete
spacetimes.
\end{remark}

\begin{remark}[Fixed and sudden  singularities]
The actual finite duration of a singular spacetime (according to the
dynamical definition), in other words the  position on the time axis
of the finite time singularity,  is in fact its maximum interval of
existence considered as a solution. This therefore may very well
depend on the initial and final conditions set in the problem. One
is therefore led to consider \emph{fixed} singularities, those which
have positions independent of the initial conditions as well as
\emph{movable}, or \emph{sudden} singularities, whose position  is
actually very sensitive to the initial conditions. An interesting
nontrivial problem is to characterize these two  types of
singularities using the geometric definitions of geodesic
incompleteness. For more details see Section \ref{Sec:sudden}.
\end{remark}

\subsection{The physical approach}
Up to around the 1960's only three dynamical theories existed in
physics: Newtonian dynamics (of particles, continua and fields), relativistic dynamics (including both
the special and the general theory) and quantum dynamics (splitting
naturally in its newtonian (mechanics) and relativistic (field
theory) parts). Since then, however, there has been a proliferation
of different dynamical theories some aiming  at rectifying one or
more of the  issues which arose in one of the three pillars, others
pointing at the other direction, that of the \emph{unification}
(eventually of relativistic and quantum dynamics). This sparkling
diversity is especially apparent in theories which include gravity,
and the following table aims at offering a broad but humble picture
of this most chaotic situation.
\begin{center}
\begin{tabular}{l|c|r}
\multicolumn{3}{c}{\emph{\textbf{Cosmologies}}}\cr\hline {\bf
{Theories of gravity}} & {\textbf{Spacetimes}} &
{\textbf{Matterfields}}\cr\hline\hline General Relativity & (Anti-)
de Sitter & Vacuum \cr Higher Derivative  & RW & Fluids \cr
Scalar-Tensor and VC & Bianchi,  & Scalar fields\cr Inflation and QC
& G\"{o}del  & $n$-form fields\cr Braneworlds   & T-NUT-M &
Phantoms, tachyons \cr\hline
\end{tabular}
\end{center}
With the advent of string theory and its subsequent generalizations
and extensions previous  dynamical theories of gravity found their
natural places under its umbrella in the form of \emph{effective}
actions and field equations and their overall position gave to the
picture a notable matter-of-fact-ness.

But it is disturbing. The whole picture is blurred with the
existence and apparent (mathematical at least) viability of all
these theories, while their structures are connected to one another
by complicated laws, duality and conformal transformations, making
it impossible to choose some and to leave more out (the existence of
the so-called string and Einstein frame formulations are good
examples of this). We don't yet know what really lies behind this
web of theories but the real question is: How is it that we can
tolerate this apparently arbitrary multiplicity? The standard answer
to this pressing question is that of course each and every one of
these possible theories in the web represent different viable
\emph{mathematical models} of reality.

The least thing one needs in this game is to become inflated with
long words. The singularity problem acquires the following simple
form in this context: Analyze the possible models (like those that
can be readily extracted from the Table above) and discover what
they have to say about the existence and nature of their spacetime
singularities. It is to be expected that the obvious proliferation
of models will lead to an equally obvious number of different kinds
of singularities. This is a good thing, it is like discovering new
kinds of life in zoology (for more on this, see \cite{me02, meL02}).

The essential mathematical tools needed to construct such
cosmological models (with the aim eventually to treat their
singularities) come from tensor analysis. One ends up with tensor
equations, invariant under arbitrary coordinate transformations, and
these describe the dynamical content of the model. Then, techniques
from the theory of dynamical systems enter the picture, and with
geometry and dynamical systems one hopes to be able to tackle the
many different subtleties which will be pertinent to the singularity
structure of the spacetimes in question. Due to its richness,
unraveling the singularities of different cosmological models is a
game that must be played with adroitness, many surprises are still
hidden from our ever expanding horizon. More on this approach will
be given in Section 3.

\subsection{The mathematical approach}
There is another way of looking at the problem of cosmological
singularities, but before we discuss this in this subsection, let us
summarize in a sentence the content of the approach taken in the
previous subsection. For the admitted variety of theories of gravity
we have to pay a price, namely, the spacetimes to-be-used are
mathematically simple, perhaps too simple, but there are at least
two good reasons that make that a legitimate approach: Even if one
does not believe that the actual models constructed this way are
true in reality,  one still has a reason to consider them for
\emph{only} in this way we have even the slightest chance to use the
available observational data (cf. \cite{wein}, p. 470).  Or, we
perhaps, like Einstein,  must elevate the \emph{homogeneity
assumption} to the status of a physical principle, cf. \cite{pee},
p. 16). Follow either of these two leads and you conveniently find
yourself walking along with the other \emph{physical cosmologists}.

Others, however, follow none. The point of view of the
\emph{mathematical cosmologist}, at least in its traditional sense,
is that, because we know so little from observations, \emph{it is
legitimate to consider arbitrary manifolds and metrics} (cf.
\cite{ycb-y02}, p.30). There is also another reason for this
legitimacy. It is quite possible that homogeneities at small scales
might change drastically the \emph{global} mode of behaviour in any
model when we follow its evolution in either early or late (proper)
times. For the treatment of homogeneous cosmological models in any
theory of gravity, this could affect the results obtained through
the physical approach quite seriously, even deadly\footnote{There
are of course certain `no-go' conjectures, the BKL conjecture being
the primary example (see, the very recent work \cite{ug} for an
account of the current status of the proof of this conjecture). For
the singularity problem, this basically says that what happens in
the most general \emph{homogeneous} model on approach to the
singularity (technically for the Bianchi IX spacetimes) describes
the generic situation, that is the approach to the singularity when
there are no symmetries.}.

The mathematical cosmologist does not have to go far to find
problems of his liking, general relativity is for the most part
adequate. Even the simplest of theories when looked from such a
perspective can have  a richness of unprecedented beauty.  However,
sooner or later the mathematical approach faces a true obstacle: The
transcendental difficulty of the equations. The Einstein  equations
are, in their most interesting form, a system of hyperbolic
evolution equations together with elliptic constraints and this
system will obviously have to be augmented by further equations
describing the evolution of the various matter fields.

However, some progress has been made and here we have the Cauchy
problem (see \cite{ycb-y02} and references therein). In this
approach, spacetimes are not known in advance, they are in fact
built up from initial data as the solutions that describe them
evolve, or develop, through the field equations from that data. The
issues of the existence and uniqueness of \emph{maximal
developments}, as well as that of the nature of singularities in
these solutions are the central objects of study here, and it is of
interest to apply these methods not only to general relativity but
also to other interesting theories of gravity formulated in the
string frame, something that has not yet been tried to any
measurable degree of completeness.

Another area in which modern mathematical methods can be of great
interest is in fact that which contains models  of interest to the
physical cosmologist. Here a more mathematical approach to models
already formulated by physicists can help in improving the rigor of
many presentations as well as unify and polish the essential
elements of many models considered in the literature. We shall have
to say more on this later.

Still another direction to the singularity problem open to the
modern mathematical cosmologist is through Lorentzian geometry. This
approach has led historically to the first singularity theorems,
those  of R. Penrose and S. W. Hawking (see \cite{o'n} for a modern
introductory account of these theorems).\footnote{There are many
results in this and related directions, see the book\cite{beem}
where  this whole approach is presented along with interesting
theorems and open problems.} Recently there has been a related body
of results, the \emph{completeness theorems} (see \cite{cbc} and
also \cite{me-k1} for an application of these results to cosmology).
These theorems aim to give sufficient conditions for a spacetime to
be geodesically complete, however, like the singularity theorems
(but unlike the dynamical mathematical approaches described above),
they require a spacetime of a \emph{known} form in order to be
applied. We shall have to say more about these theorems in Section
4.

\section{Geometric asymptotics for exact models}
A large part of the literature\footnote{To do full justice to the
enormous number of published papers in the many different areas of
this subject can lead to a project in
itself, something that this humble review never intended to do (a
comprehensive \emph{Bibliography of Papers on Cosmological
Singularities} is currently under preparation). We
apologize in advance to many readers to whom the choice of
references included here, and more importantly their omissions,
seems arbitrary. The present author is aware of this issue only too
well!} on cosmological singularities is directly concerned with the
problem of tracing out the exact, early or late time, asymptotic
behaviour of various cosmological models. In all these studies, a
cosmology (in the sense of the previous Section) is given and the
problem  is then to determine the nature of the spacetime and of the
various fields in the neighborhood of a past or future singularity.
This is basically an \emph{asymptotic problem} since the form of the
spacetime is known in advance but the precise asymptotic relations
of the unknown quantities present in the cosmology are to be
determined. In this Section, we present an outline of a new (or more
precisely cleaned-up) method, called hereafter \emph{the method of
asymptotic splittings} (cf. \cite{CB} for a more detailed
presentation), which enables us to precisely determined such
asymptotic relations. The utility of the method of asymptotic
splittings to analyze cosmological singularities is then shown in a
number of recently examined cases (with references to the relevant
papers). We also add some comments and remarks about the issue of
fixed vs. spontaneous (or sudden) singularities, the latter having
recently gained some attention, in a slightly different context,
through the work of J. D. Barrow and others (see \cite{ba1, ba2,
me-k1} and refs. therein).

\subsection{Method of asymptotic splittings}
In this subsection, we follow \cite{CB} closely and this reference is
to be consulted for notation, explanations, and proofs of any stated
result. The general setting is a vector field
$\mathbf{f}:\mathcal{M}^{n}\rightarrow \mathcal{TM}^{n}$ and the
associated dynamical system defined by $\mathbf{f}$ on the manifold
$\mathcal{M}^{n}$,
\be\dot{\mathbf{x}}=\mathbf{f}(\mathbf{x}%
),\ee with $(\cdot )\equiv d/dt$. Depending on the number of
arbitrary constants, a solution of our dynamical system can be
\emph{general, particular} or \emph{exact}. Any  solution, however,
can develop
\emph{finite-time singularities}, that is instances where a solution $%
\mathbf{x}(t;c_{1},\cdots ,c_{k}),\,k\leq n$, misbehaves at a finite value $%
t_{\ast }$ of the time $t$. This is made more precise as follows. We
say that the system $\dot{\mathbf{x}}=\mathbf{f}(\mathbf{x})$
(equivalently, the vector
field $\mathbf{f}$) has a \emph{finite-time singularity} if there exists a $%
t_{\ast }\in \mathbb{R}$ and a $\mathbf{x}_{0}\in \mathcal{M}^{n}$
such that for all $M\in \mathbb{R}$ there exists an $\delta >0$ such
that
\begin{equation}
||\mathbf{x}(t;\mathbf{x}_{0})||_{L^{p}}>M,  \label{sing1}
\end{equation}%
for $|t-t_{\ast }|<\delta $. Here $\mathbf{x}:(0,b)\rightarrow \mathcal{M}%
^{n}$, $\mathbf{x}_{0}=\mathbf{x}(t_{0})$ for some $t_{0}\in (0,b)$, and $%
||\cdot ||_{L^{p}}$ is any ${L^{p}}$ norm. We say that the vector
field has
a \emph{future} (resp. \emph{past}) singularity if $t_{\ast }>t_{0}$ (resp. $%
t_{\ast }<t_{0}$). Note also, that $t_{0}$ is an arbitrary point in
the domain $(0,b)$ and may be taken to mean `now'. We often write
\begin{equation}
\lim_{t\rightarrow t_{\ast
}}||\mathbf{x}(t;\mathbf{x}_{0})||_{L^{p}}=\infty ,  \label{sing2}
\end{equation}%
to denote a finite-time singularity at $t_{\ast }$. Our basic
problem then is to find the structure of the set of points
$\mathbf{x}_{0}$ in $\mathcal{M}^n$ such that, when evolved through
the dynamical system defined by the vector field, the integral curve
of $\mathbf{f}$ passing through a point in that set satisfies
property (\ref{sing2}).

There are two kinds of finite time singularities that a nonlinear
dynamical system can possess, \emph{fixed} and \emph{movable}
(sudden, or spontaneous in other terminology is  often used for a
movable singularity). A singularity is \emph{fixed} if it is a
singularity of $\mathbf{x}(t;\mathbf{C})$ for all $\mathbf{C}$;
otherwise, we say it is a \emph{movable singularity}. Our basic
problem is then: What can a vector field do, or equivalently, how do
the solutions of the associated dynamical system behave in the
neighborhood of a finite time singularity? Assume that we are given
a vector field and suppose that at some point, $t_{\ast }$, a system
of integral curves, corresponding to a particular or a general
solution, has a (future or past) finite-time singularity in the
sense of definition (\ref{sing1}).  The vector field (or its
integral curves, solutions of the dynamical system defined by
$\mathbf{C}$) can basically do two things sufficiently close to the
finite-time singularity, namely, it can either show some dominant
feature or not. In the latter case, the integral curves can `spiral'
in some way around the singularity \emph{ad infinitum} so that
(\ref{sing1}) is satisfied and the dynamics is totally controlled by
subdominant  terms, whereas in the former case solutions share a
distinctly dominant behaviour on approach to the singularity at
$t_{\ast }$ determined by the most nonlinear terms.

Our approach to this problem is an asymptotic one. We decompose, or
split,  the vector field into simpler, component vector fields and
examine whether the most nonlinear one of these shows a dominant
behaviour while the rest become subdominant in some exact sense. We
then built a system of integral curves corresponding, where
feasible, to the general solution and sharing exactly its
characteristics in a sufficiently small neighborhood of the
finite-time singularity.  In this way we are led to a general
procedure to uncover the nature of singularities by constructing
series expansion representations of particular or general solutions
of dynamical systems in suitable neighborhoods of their finite-time
singularities.

This \emph{method of asymptotic splittings} \cite{CB} consists of
building splittings of vector fields that are valid asymptotically
and trace the dominant behaviour of the vector field near the
singularity. A resulting series expansion connected to a particular
\emph{dominant balance} helps to decide whether or not the arrived
solution is a general one and to spot the exact positions of the
arbitrary constants as well as their role in deciding about the
nature of the time singularity. To apply the method of asymptotic
splittings to a given dynamical system so as to discover the nature
of its solutions near the time singularities, we must follow a
number of steps:
\begin{enumerate}
\item Write the system of equations in the form of a dynamical system $\dot{%
\mathbf{x}}=\mathbf{f}(\mathbf{x})$ with
$\mathbf{x}=(x_1,\cdots,x_n)$, and
identify the vector field $\mathbf{f}(\mathbf{x})=(f_1(\mathbf{x}%
),\cdots,f_n(\mathbf{x}))$.

\item Find all the different weight-homogeneous decompositions of the
system, that is the splittings of the form
\[
\mathbf{f}=\mathbf{f}^{\,(0)}+\mathbf{f}^{\,(1)}+\cdots
+\mathbf{f}^{\,(k)},
\]%
and choose one of these splittings to start the procedure.

\item Substitute the scale-invariant solution
\[
\mathbf{x}^{(0)}(\tau )=\mathbf{a}\tau ^{\mathbf{p}},
\]%
into the equation $\dot{\mathbf{x}}=\mathbf{f}^{(0)}$. Study the
resulting algebraic systems, and find all dominant balances
$(\mathbf{a},\mathbf{p})$ together with their orders.

\item Identify the non-dominant exponents, that is the positive numbers
$q^{(j)},\, j=1,\cdots,k$, such that
 \[
\tau
^{q^{(j)}}\sim\frac{\textbf{f}^{\,\,\textrm{sub},\,{(j)}}(\tau^{\bf
p})} {\tau^{\mathbf{p}-\mathbf{1}}}\rightarrow 0.
\]

\item Construct the K-matrix $\mathcal{K}$:
\[
\mathbf{f}^{(0)}\rightarrow D\mathbf{f}^{(0)}\rightarrow D\mathbf{f}^{(0)}(%
\mathbf{a})\rightarrow
D\mathbf{f}^{(0)}(\mathbf{a})-\textrm{diag}\,\mathbf{p} .
\]

\item Compute the spectrum of $\mathcal{K}$,
\[
\textrm{spec}(\mathcal{K})=(-1,\rho _{2},\cdots ,\rho _{n}).
\]%
Is $\mathcal{K}$ semi-simple? Are the balances hyperbolic?

\item Find the eigenvectors $\mathbf{v}^{(i)}$ of $\mathcal{K}$.

\item Identify $s$ as the multiplicative inverse of the least common
multiple of all the subdominant exponents and positive K-exponents.

\item \label{puiseux} Substitute the Puiseux series
\[
x_{i}=\sum_{j=0}^{\infty }c_{ji}\,\tau ^{p_{i}+\frac{j}{s}}
\]%
into the original system.

\item Identify the polynomials $\mathbf{P}_{j}$ and solve for the final
recursion relations which give the unknown coefficients
$\mathbf{c}_{j}$.

\item Check the compatibility conditions at the K-exponents,
\[
\mathbf{v}^{\top}_\rho\cdot\mathbf{P}_\rho=0,\quad \textrm{for each eigenvalue}%
\,\,\rho.
\]

\item If the Puiseux series is valid, then the method is concluded for this
particular splitting. Otherwise, if compatibility conditions are
violated at the eigenvalue $\rho ^{\ast }$, restart from step
\ref{puiseux} by substituting the logarithmic series
\[\label{main4}
\mathbf{x}=\tau^{{\bf
p}}\left(\mathbf{a}+\sum_{i=1}^{\infty}\sum_{j=1}^{\infty}
\mathbf{c}_{ij}\tau^{i/s}(\tau^\rho\log\tau)^{j/s}\right), \]

\item Get coefficient at order $\rho ^{\ast }$. Write down the final
expansion with terms up to order $\rho ^{\ast }$.

\item Verify that compatibility at $\rho^*$ is now satisfied.

\item Repeat whole procedure for each of the other possible decompositions.
\end{enumerate}
Following the above steps even up to that of calculating a dominant
balance in one particular decomposition (that is up to step 4), can
be very useful since it offers one particular possible asymptotic
behaviour of the system near the time singularity. In this respect,
the whole method expounded here is truly generic since it helps to
decide the generality of any behaviour found in an exact solution --
that is, how many arbitrary constants there are in the final
solution that shares that behaviour (particular or general
solution). It is rare that a Puiseux series is inadequate to
describe the dynamics (semi-simplicity of K), but in such  cases one
must resort to the more complex logarithmic solutions.

\subsection{Example: Gauss-Bonnet cosmologies}
There are many interesting systems that could be seen under the
magnifying glass of the method of asymptotic splittings, cf. the
systems in \cite{CB,cm,cak,ct1}. Here we present a short summary of
some of the results of our very recent work \cite{ct2} in
collaboration with A. Tsokaros, which is to be followed for
motivation and further explanations. The starting point is the
action
\be\mathcal{S}=\int_{\mathcal{M}^4}\mathcal{L}_{\textrm{total}}d\mu_{g},\quad
\quad d\mu_{g}=\sqrt{-g} d\Omega,\ee where
$\mathcal{L}_{\textrm{total}}$ is the lagrangian density of the
general quadratic gravity theory given in the form\footnote{the
conventions for the metric and the Riemann tensor are those of
\cite{ll}.}
$\mathcal{L}_{\textrm{total}}=\mathcal{L}(R)+\mathcal{L}_{\textrm{matter}}$,
with \be \mathcal{L}(R)=R + BR^2 + C\textrm{Ric}^2 +
D\textrm{Riem}^2 , \label{eq:lagra} \ee where $B,C,D$ are constants.
Since in four dimensions we have the Gauss-Bonnet  identity, \be \GD
\int_{\mathcal{M}^4}R^2_{GB}d\mu_{g}=0,\quad R^2_{GB}=R^2 -
4\textrm{Ric}^2 + \textrm{Riem}^2, \label{eq:gentity} \ee in the
derivation of the field equations through variation  of the action
associated with (\ref{eq:lagra}), only terms up to $\textrm{Ric}^2$
will matter. Below we focus exclusively in spatially flat universes
of the form \be ds^2=dt^2-b(t)^2(dx^2+dy^2+dz^2),
\label{eq:flatun}\ee which are radiation dominated ($P=\GR/3$), and
use only the 00-component of the field equations, this being the
following equation satisfied by the scale factor $b(t)$: \be
\frac{\dot{b}^2}{b^2}-\GK\left[2\: \frac{\dddot{b}\:\dot{b}}{b^2} +
2\:\frac{\ddot{b}\dot{b}^2}{b^3}-\frac{\ddot{b}^2}{b^2} - 3\:
\frac{\dot{b}^4}{b^4} \right] - \frac{b_{1}^2}{b^4} =0 ,
\label{eq:beq} \ee where $b_{1}$ is a constant defined by \be
\frac{8\pi G \rho}{3c^4}=\frac{b_{1}^2}{b^4},\quad
(\textrm{from}\,\,\nabla_{i}T^{i0}=0). \ee Note that the Friedmann
solution $\sqrt{2b_1 t}$ of general relativity satisfies the above
equation.

To apply the method of asymptotic splittings, we set $b=x$,
$\dot{b}=y$ and $\ddot{b}=z$,  and then Eq. (\ref{eq:beq}) can be
written as a dynamical system of the following form:
$\mathbf{\dot{x}}=\mathbf{f}(\mathbf{x})$, $\mathbf{x}=(x,y,z)$:
\begin{equation}
\label{eq:ds} \dot{x} = y,\:\:\:\:\: \dot{y} = z,\:\:\:\:\: \dot{z}
= \frac{y}{2\GK} - \frac{b_{1}^2}{2\GK yx^2} - \frac{yz}{x} +
\frac{z^2}{2y} + \frac{3y^3}{2x^2}.
\end{equation}
Then we look for the possible weight-homogeneous decompositions. The
 most interesting one has \emph{dominant part}
\be
\mathbf{f}^{(0)}=\left(y,z,\frac{z^2}{2y}-\frac{zy}{x}+\frac{3y^3}{2x^2}
\right), \label{eq:f0} \ee while the subdominant part reads \be
\mathbf{f}^{\,\textrm{sub}}=\left(0,0,-\frac{b_{1}^{2}}{2\GK
yx^2}+\frac{y}{2\GK}  \right), \ee with $\mathbf{f}=\mathbf{f}^{(0)}
+ \mathbf{f}^{\,\textrm{sub}}$. The dominant balance for this
decomposition turns out to be the unique (full order) form \be
(\mathbf{a},\mathbf{p}) = \left( \left(
\GA,\frac{\GA}{2},-\frac{\GA}{4}\right),\:
\left(\frac{1}{2},-\frac{1}{2},-\frac{3}{2} \right) \right),
\label{eq:domibal} \ee where $\GA$ is an arbitrary constant.

Next we turn to an examination of the form and properties of the
K-matrix. The Kowalevskaya exponents for this particular
decomposition, eigenvalues of the matrix
$\mathcal{K}=D\mathbf{f}(\mathbf{a})-\textrm{diag}(\mathbf{p})$, are
$\{-1,0,3/2\}$ with corresponding eigenvectors
$\{(4,-2,3), (4,2,-1),\\ (1,2,2)\}$. (The arbitrariness of the
coefficient $\GA$ in the dominant balance reflects the fact that one
of the dominant exponents is zero with multiplicity one.) After
following the other steps of the method we finally arrive at the
following series expansion representation of the behaviour of the
system in the neighborhood of the initial collapse singularity: \be
x(t) = \GA \:\: (t-t_{0})^{1/2} + c_{13} \:\: (t-t_{0})^{2} +
\displaystyle \frac{\GA^4-4b_{1}^2}{24\GK\GA^3} \:\:
                    (t-t_{0})^{5/2} + \cdots .
\label{eq:sol} \ee The series expansions for $y(t)$ and $z(t)$ are
given by  the first and second time derivatives of the above
expressions respectively.

There are several comments to be made for this solution. First, the
series (\ref{eq:sol}) has three arbitrary constants, $\GA, c_{13},
t_0$ (the last corresponding to the arbitrary position of the
singularity) and is therefore a local expansion of the
\emph{general} solution around the movable singularity $t_0$.
Secondly, our solutions are stable in the neighborhood of the
singularity in the sense that  all flat, radiation solutions of the
quadratic gravity theory considered here are Friedmann-like
regardless of the sign of the $R^2$ coefficient. This implies that
there are no bounce solutions of the model analyzed here. It remains
an open problem to find a physically reasonable modification of this
model with stable, bouncing solutions valid at early times.

\section{Differential geometric issues}
There are many instances where one does not wish to put precise
assumptions on  the form of the cosmological spacetime but leave it
unspecified instead. In such circumstances it is important to be
able to find general conditions and criteria under which a
`universe' represented in such a generic way will develop a
physical, past or future, singularity in a finite time. In fact, it
turns out that one can formulate precise criteria for such purposes
and even begin to classify the possible mathematical and physical
singularities of this generic situation.

In this section we review some attempts to tackle this problem,
beginning with the (statement of the) time-honored singularity
theorems, results predicting the general existence of spacetime
singularities, first formulated in the sixties. These results
utilized in their proof some of the most beautiful concepts and
techniques from differential topology and Riemannian geometry, but
since then the general theory has remained somewhat dormant and
perhaps wanting. Recently, it has been possible to revive interest
in this whole area and new results predicting the existence of
complete as well as singular spacetimes have been made possible,
mainly due to the efforts of Y. Choquet-Bruhat. An adaptation of
these ideas to simple cosmological solutions reveals structures
which are perhaps difficult to unearth by more traditional methods
and analysis.

\subsection{Completeness and  singularities}
The geometric criteria to be developed in this subsection for a
spacetime to be either geodesically complete or incomplete will be
based on certain properties of its spacelike submanifolds. Below we
focus exclusively on spacelike \emph{hypersurfaces}, although
submanifolds of higher codimension also play a significant role for
such purposes (the most well-known example of the latter situation
is the notion of a closed trapped surface).

Consider a spacetime of the form $(\mathcal{M},g)$ with
$\mathcal{M}=\Sigma \times \mathcal{I},\;$ $\mathcal{I}$ being an
interval in $\mathbb{R}$ (for simplicity we can take it to be the
whole real line) and $(\Sigma,g_t)$  a smooth, spacelike submanifold
of dimension $n$, in which the smooth, $(n+1)$--dimensional,
Lorentzian metric $g$ splits as follows:
\begin{equation}
g\equiv -N^{2}(\theta ^{0})^{2}+g_t,\quad g_t\equiv g_{ij}\;\theta
^{i}\theta ^{j},\quad \theta ^{0}=dt,\quad \theta ^{i}\equiv
dx^{i}+\beta ^{i}dt. \label{2.1}
\end{equation}
Here $N=N(t,x^{i})$ denotes the \emph{lapse function}, $\beta
^{i}(t,x^{j})$ is the \emph{shift vector field} and all spatial
slices $\Sigma_{t}\,(=\Sigma\times \{t\})$ are spacelike
hypersurfaces endowed with the time-dependent spatial metric
$g_{t}\equiv g_{ij}dx^{i}dx^{j}$.  We assume that our spacetime
$(\mathcal{V},g)$ is \emph{regularly sliced}, that is the lapse
function is bounded, the shift vector field is uniformly bounded and
the spatial metric is itself uniformly bounded below, cf.
\cite{cbc,co04}.

There are two kinds of information about the slices $\Sigma_{t}$
that play a role in what follows: topological and geometric. The
topological part of the theory to-be-developed starts with a
connection of the completeness of $\Sigma_{t}$ with the global
hyperbolicity of $(\mathcal{M},g)$, as in the following result from
\cite{cbc,co04} which may be consulted for the proof.
\begin{theorem}
Let  $(\mathcal{M},g)$ be a regularly  sliced spacetime. Then the
following are equivalent:
\begin{enumerate}
\item $(\Sigma_{0},g_{0} )$ is a $g_0$-complete Riemannian manifold
\item The spacetime $(\mathcal{M},g)$ is globally hyperbolic
\end{enumerate}
\end{theorem}
We thus see that our ability to successfully exercise classical
determinism in a general spacetime is intimately connected to the
topological property of geodesic completeness, not of the whole of
spacetime, but of each one of its spacelike slices (considered as
Riemannian manifolds, so that, in particular the Hopf-Rinow theorem
is valid). Unfortunately to say something about the geodesic
completeness of the \emph{whole} of our general spacetime
$(\mathcal{M},g)$ is not possible by using solely topological pieces
of information as the one given above (unless $(\mathcal{M},g)$ is
trivial in some sense, see\cite{co04}).

We need to know in addition more about the geometry of
$(\mathcal{M},g)$. What we specifically need to know in order to
state the following singularity as well as completeness theorems, is
some \emph{submanifold geometry} (cf., for example, \cite{o'n} chap.
4 and parts of chap. 10). For both kinds of theorem, the notion the
\emph{shape tensor} (or second fundamental form) of the slice
$\Sigma_{t}$ becomes vital. Consider the mean curvature vector field
$H$ of $\Sigma_{t}$, denote by $\bar{\nabla}$ the induced connection
on $\Sigma_{t}$, and let $K=\textrm{nor}\,\bar{\nabla}$ be its shape
tensor which is defined on pairs of vector fields of $\Sigma_{t}$.

The trace and norm of $K$ play a fundamental role. For a unit future
pointing vector field $U$ normal to $\Sigma_{t}$, we define the
\emph{convergence} of $\Sigma_{t}$ to be the real-valued function
$\theta$ on the normal bundle $N\Sigma_{t}$ \be\theta=\left\langle
U,H\right\rangle =\frac{1}{n-1}\,\textrm{trace}\, K. \ee A natural
interpretation of $\theta$ is to say that when  $\theta$ is positive
the spacelike slice is bent inward so that  its outward pointing
normals converge, whereas when it is negative we have the reverse
situation. One may easily then imagine a spatial slice such that
$\theta$ is positive \emph{everywhere} on it: we will picture it as
a convex spacelike arc of infinite extent.

The only other  piece of geometric information we shall need is
this: the $g_t$ norms of $K$ as well as of the spatial gradient of
the lapse function, namely, $|K|_{\mathbf{g}_t}$ and $|\nabla
N|_{\mathbf{g}_t}$. These are the natural objects to consider and
estimate in an effort to see how large or small they can become and
so to develop a feeling on how the slice itself evolves with proper
time.

We are now ready to state a basic singularity theorem and a
completeness theorem to get a taste of this kind of result (for more
on this see the references above). The first result is Hawking's
singularity theorem.
\begin{theorem}If:
\begin{itemize}
\item {$\textrm{Ric}(X,X)\geq 0$} for all causal vector fields {$X$}
of  the spacetime {$(\mathcal{M},g)$}

\item  {$\theta\geq C> 0$}, {everywhere} on the
Cauchy slice {$\Sigma$},
\end{itemize}
then no future-directed causal curve from {$\Sigma$} can have length
greater than {$1/C$}.
\end{theorem}
This theorem continues to  hold if in the place of the words `Cauchy
slice' one substitutes the weaker `compact slice', although in this
case  the proof is somewhat more involved (cf. \cite{o'n} pp.
431-3). The proof is by contradiction: One cannot have both a
complete spacetime \emph{and} a Cauchy (or compact) slice with a
positive convergence \emph{everywhere}. That is unlike the case
where the convergence were positive only locally, then nothing bad
would happen. (This is the case, for instance, when one considers a
convex spacelike arc \emph{of finite length} with positive
convergence in two-dimensional Minkowski space: its normals are
converging, they meet at some point to the future of the arc but
there is no singularity, spacetime is geodesically complete.)
Therefore this theorem predicts that under this situation spacetime
must be geodesically incomplete, that is singular, but in a very
special way: the whole of spacetime will eventually fold around into
itself!

On the brighter side, we have the following completeness result
\cite{cbc}.
\begin{theorem}
If
\begin{itemize}
\item $(\mathcal{M},g)$ is a globally hyperbolic, regularly sliced spacetime
\item for each finite {$t_{1},$ $|\nabla N|_{\mathbf{g}_t}$} and
{$|K|_{\mathbf{g}_t}$}  are {integrable} functions  on
{$[t_{1},+\infty )$},
 \end{itemize}
then {$(\mathcal{M},g)$} is future causally geodesically complete.
\end{theorem}
The proof of this result is achieved by showing that all
future-directed geodesics have an infinite length, and it is done by
analyzing properties of the geodesic equation. For various
applications of this theorem to cosmological spacetimes, see
\cite{me-k1,me-kl1a}.

\subsection{Fixed and sudden singularities}\label{Sec:sudden}
We pause in this subsection to discuss, in passing, some qualitative
differences between the two possible \emph{dynamical} kinds of
finite time singularities inherent in the method of asymptotic
splittings, namely, fixed and movable (or sudden) singularities.
Both types of singularities are of course in accordance with the
assumptions of the singularity theorems, although in discussions of
the latter the distinction is never made explicit. (For
concreteness, the reader if he so wishes, may take a simple
cosmological metric, for instance one from the Robertson-Walker
family, which has only one unknown function, the scale factor $a$.)
We assume that we have a situation where the assumptions of the
singularity theorems are valid and the model has a true, finite time
singularity. Our problem currently is to obtain conditions under
which this singularity is fixed or movable.

Suppose we have a singly infinite family of geodesics in our
spacetime. These geodesics form a 2-dimensional geodesic surface
parametrized by
 \be
x^r=x^r(u,v),
 \ee
and one understands that the geodesics defined in this way are the
parametric lines of $u$ in the geodesic surface. We denote by
$p^r=\partial x^r  /\partial u$ the unit tangent vector field to the
geodesics, and by $\eta^r=(\partial x^r  /\partial v)dv$ the
connecting, infinitesimal vector field between two adjacent
geodesics of the family, also called a \emph{Jacobi field}. These
two vectors $p^r, \eta^r$ satisfy two fundamental equations, namely
the geodesic equation and the Jacobi equation (or equation of
geodesic deviation) respectively. Both  equations are very familiar
and play a basic role in any discussion of geodesic incompleteness.
They read: \be\label{g-eq}
\dot{p}^r+\Gamma^r_{mn}p^mp^n=0,\quad\textrm{Geodesic Equation}, \ee
and \be\label{j-eq} \ddot{\eta}^r+R^r_{smn}p^s\eta^m
p^n=0,\quad\textrm{Jacobi Equation}. \ee Here, differentiation is
meant with respect to arc length (wherever it makes sense). Both
equations are nonlinear but their main difference is this: Whereas
the equation of geodesics is not a tensor equation  but a
frame-dependent one and so any conclusions using it may change under
coordinate transformations, the Jacobi equation is a tensorial
equation and anything we say using it will be valid in an invariant
geometric way.

For instance, although the existence of conjugate points (two events
each having $\eta =0$) on two adjacent geodesics follows from an
analysis of the Jacobi equation and is true in any system once it is
established, properties of solutions of the geodesic equation are
not so insensitive.

Suppose we have a Robertson-Walker metric with a true singularity,
as, for instance, the initial big bang (collapse) singularity in the
standard cosmological models, cf.\cite{ll}, p. 366. The actual
spacetime point, location of the singularity, is then a conjugate
point in the sense that there is a non-zero Jacobi field between any
later point and the big bang that becomes zero there. Timelike
geodesics that end at the big bang (which in this case is an
all-encompassing singularity) are still solutions of the geodesic
equation (\ref{g-eq}). From standard expressions, the Christoffel
symbols diverge at the singularity, in other words the
\emph{coefficients} in the geodesic equation (\ref{g-eq}) are
diverging. This means that the standard big bang singularity is a
\emph{fixed} singularity in the terminology of the previous Section.
It does not depend on the initial conditions (which may be taken to
be at any later time),  in this case the singularities of the
dynamical system are read off from the singularities of its
coefficients.

A completely different situation may however arise, and this is
reminiscent with what happens to  solutions of the very simple
nonlinear  equation $\dot x =x^2$ with initial condition $x(0)=x_0$.
There is the solution $x(t)=x_0/(1-x_0 t)$ and this is defined only
on $(-\infty, 1/x_0)$, when $x_0>0$, on the whole of $\mathbb{R}$
for $x_0=0$, and on $(1/x_0,\infty)$ for $x_0<0$. We see that the
solution ceases to exist after time $1/x_0$. The coefficient in the
equation is, however, regular (in fact equals one). So this equation
has singularities whose position depends on the initial conditions
and they `move' accordingly with the initial condition, but the
coefficients of the equation don't `see' them. This is the simplest
example of a movable (or sudden) singularity.

We believe that in general relativity we can in principle have a
similar situation. For, if a geodesic defined on an interval $(a,b)$
is, say, future incomplete (as that would follow from a tensorial
analysis of the invariant Jacobi equation), then it cannot be
extended after $b$ to the whole of the real line, but this $b$ may
or may not depend on the initial conditions chosen, making the
singularity either fixed or movable. This is only to be expected
because the geodesic equation (\ref{g-eq}) is nonlinear and so
cannot have singularities of only the fixed kind (that is those that
are read off from the coefficients of the equation). Singularities
in the traditional form of g-incompleteness are usually thought of
in the literature only as fixed singularities, no dependence of the
singularity on the `initial' (later for a past, earlier for a future
movable singularity) conditions.

\subsection{Bel-Robinson energy}
It is clear from what has been said earlier that there are several
different, inequivalent ways to characterize cosmological
singularities. First, there is the traditional way, through
differential-geometric methods, as explained in the standard text on
this approach \cite{he}. This set of methods uses a global approach
to spacetimes with an emphasis on the causal, topological aspects of
singularities. But there is more to characterizing singularities
than this, and using \emph{dynamics} one may arrive at a
complementary set of tools for such purposes. We saw how, using the
method of asymptotic splittings, we can have quantitative
information about a found singularity and form a clear picture on
the different modes of approach through series expansions of the key
unknown functions pertinent to the given problem. Other relevant
approaches based on dynamical systems are also of great use, cf.
\cite{we}. A third approach is to use the contrapositive of the
completeness theorem discussed earlier and end up with
\emph{necessary conditions} for singularities. This approach to the
nature of cosmological singularities was taken up in
\cite{me-k1,me-kl1a} for the simple case of Robertson-Walker
cosmologies. The result is that we can arrive at an interesting
trichotomy of the singularities in various Friedman cosmologies by
using the two main functions of the problem, namely, the Hubble
expansion rate $H$ and the scale factor $a$.

To the three main approaches to singularities another must be added.
In none of these does the presence of matter fields play a role, at
least directly, and the following way to classify cosmological
singularities uses exactly this missing piece of information: It is
based on the introduction of an invariant geometric quantity, the
\emph{Bel-Robinson energy} which takes into account precisely those
features of the problem, related to the matter contribution, in
which models still differ near the time singularity while having
similar behaviours of $a$ and $H$. In this way, we arrive at a
complete classification of the possible cosmological singularities
in the isotropic case (for more details see \cite{k2,k3,k4}).

The \emph{Bel-Robinson energy at time $t$} is given by \be
\mathcal{B}(t)=\frac{1}{2}\int_{\mathcal{M}_t}\left(|E|^{2}+|D|^{2}+
|B|^{2}+|H|^{2}\right)d\mu_{\bar{g}_t}, \ee where by
$|X|^{2}=g^{ij}g^{kl}X_{ik}X_{jl}$ we denote the spatial norm of the
$2$-covariant tensor $X$, and the time-dependent space electric and
magnetic tensors comprising a so-called \emph{Bianchi field} are
given in tensorial notation by \be E_{ij}=R^{0}_{i0j},\,
D_{ij}=\frac{1}{4}\eta_{ihk}\eta_{jlm}R^{hklm},\,
H_{ij}=\frac{1}{2}N^{-1}\eta_{ihk}R^{hk}_{0j},\,
B_{ji}=\frac{1}{2}N^{-1}\eta_{ihk}R^{hk}_{0j}. \ee For any
Robertson-Walker spacetime, $ ds^2=-dt^2+a^2(t )d\sigma ^2$, the
norms of the magnetic parts, $|H|, |B|$, are identically zero while
$|E|$ and $|D|$, the norms of the electric parts, reduce to the
forms \be |E|^{2}=3\left(\ddot{a}/{a}\right)^{2}
\quad\textrm{and}\quad
|D|^{2}=3\left(\left(\dot{a}/{a}\right)^{2}+k/{a^{2}}\right)^{2}.
\ee Therefore the Bel-Robinson energy becomes \be
\mathcal{B}(t)=\frac{C}{2}\left(|E|^{2}+|D|^{2}\right), \ee where
$C$ is the constant volume of (or \emph{in} in the case of a
non-compact space) the 3-dimensional slice at time $t$. Thus we find
that \be \mathcal{B}(t)\sim k_u^2(t)+k_{\sigma}^2(t), \ee where
$k_u,\,k_{\sigma}$ are the \emph{principal sectional curvatures} of
this space.  Using this result, it is not difficult to show the
following theorem concerning the completeness of Friedmann
cosmologies following directly from bounds on their Bel-Robinson
energy, and an associated minimum radius.
\begin{theorem}
A spatially closed, expanding at time {$t_*$}, Friedmann universe
that satisfies {$\gamma<\mathcal{B}(t)<\Gamma$}, where $\gamma,
\Gamma$ are constants,  is causally geodesically complete. Further,
there is a minimum radius, $a_{\textrm{min}}>\Delta^{-1/2}$,
$\Delta$ depending on $\Gamma, $ and these universes are eternally
accelerating {($\ddot{a}>0$)}.
\end{theorem}
The ensuing classification meeting all the different criteria
discussed previously results from the possible combinations of the
three main functions in the problem, namely, the scale factor $a$,
the Hubble expansion rate $H$ and the Bel Robinson energy
$\mathcal{B}.$ The singularity types will by necessity entail a
possible blow up in the functions $|E|$, $|D|$. If we suppose that
the model has a finite time singularity at time $t=t_s$, then the
possible behaviours of the functions in the triplet
$\left(H,a,(|E|,|D|)\right)$ are as follows:
\begin{description}
\item [$S_{1}$] $H$ non-integrable on $[t_{1},t]$ for every $t>t_{1}$

\item [$S_{2}$] $H\rightarrow\infty$  at $t_{s}>t_{1}$

\item [$S_{3}$] $H$ otherwise pathological

\item [$N_{1}$] $a\rightarrow 0$

\item [$N_{2}$] $a\rightarrow a_{s}\neq 0$

\item [$N_{3}$] $a\rightarrow \infty$

\item [$B_{1}$] $|E|\rightarrow\infty,\, |D|\rightarrow \infty$

\item [$B_{2}$] $|E|<\infty,\, |D|\rightarrow \infty $

\item [$B_{3}$] $|E|\rightarrow\infty,\, |D|< \infty $

\item [$B_{4}$] $|E|<\infty,\, |D|< \infty $.
\end{description}
The nature of a prescribed  singularity is thus described completely
by specifying the components in a triplet of the form
\[(S_{i},N_{j},B_{l}),\] with the indices $i,j,l$ taking their
respective values as above. Here   category $S$ monitors the
asymptotic behaviour of the expansion rate, closely related to the
extrinsic curvature of the spatial slices, $N$ that of the scale
factor, describing in a sense what the whole of space eventually
does, while $B$ describes how the matter fields contribute to the
evolution of the geometry on approach to the singularity. We know
from the completeness theorems that all these quantities need to be
uniformly bounded to produce geodesically complete universes.
Outside complete universes,  the whole situation can be very
complicated and the classification above exploits what can happen in
such a case when we consider the relatively simple geometry of
isotropic cosmologies.

\end{document}